\documentclass[aps,prb,superscriptaddress,floatfix,twocolumn]{revtex4-1}
\usepackage[pdftex]{graphicx} 
\usepackage[update]{epstopdf}
\usepackage{amssymb}
\usepackage{amsmath}
\usepackage{braket}
\usepackage{color}
\usepackage{textcomp}
\usepackage{esint}
\usepackage[utf8]{inputenc}
\makeatletter
\def\input@path{{./figures/}}
\makeatother
\usepackage{graphicx}
\usepackage{caption, subcaption}
\usepackage[export]{adjustbox}
\usepackage[colorlinks = true,
linkcolor = blue,
urlcolor  = blue,
citecolor = red,
anchorcolor = blue]{hyperref}
\usepackage{amsfonts}
\usepackage{bm}
\usepackage{multirow}
\usepackage{color}
\definecolor{dgreen}{rgb}{0,0.7,0}

\usepackage{tikz}
\usetikzlibrary{shapes,shapes.multipart}

\begin{document}
 \title{Spatio-temporal spread of perturbations in power-law models at low temperatures: Exact results for OTOC}
\author{Bhanu Kiran S.}
\address{International Centre for Theoretical Sciences, Tata Institute of Fundamental Research, Bengaluru -- 560089, India}
\author{ David A. Huse}
\address{Department of Physics, Princeton University, Princeton, NJ 08544, USA}
\author{Manas Kulkarni}
\address{International Centre for Theoretical Sciences, Tata Institute of Fundamental Research, Bengaluru -- 560089, India}
\date{\today}

\begin{abstract}
	We present exact results for the classical version of the Out-of-Time-Order Commutator (OTOC) for a family of power-law models consisting of $N$ particles in one dimension and confined by
an external harmonic potential. These particles are interacting via power-law interaction of the form $\propto \sum_{\substack{i, j=1 (i\neq j)}}^N|x_i-x_j|^{-k}$ $\forall$ $k>1$ where $x_i$ is the position of the $i^\text{th}$ particle. We present numerical results for the OTOC for finite $N$ at low temperatures and short enough times so that the system is well approximated by the linearized dynamics around the many body ground state. In the large-$N$ limit, we compute the ground-state dispersion relation in the absence of external harmonic potential exactly and use it to arrive at analytical results for OTOC. We find excellent agreement between our analytical results and the numerics. We further obtain analytical results in the limit where only linear and leading nonlinear (in momentum) terms in the dispersion relation are included. The resulting OTOC is in agreement with numerics in the vicinity of the edge of the ``light cone''.  We find remarkably distinct features in OTOC below and above $k=3$ in terms of going from non-Airy behaviour ($1<k<3$) to an Airy universality class ($k>3$). We present certain additional rich features for the case $k=2$ that stem from the underlying integrability of the Calogero-Moser model. We present a field theory approach that also assists in understanding certain aspects of OTOC such as the sound speed. Our findings are a step forward towards a more general understanding of the spatio-temporal spread of perturbations in long-range interacting systems. 

\end{abstract}
\maketitle

\textit{Introduction:}  Collective behaviour of many particle systems far-from-equilibrium has been a central issue of interest. In particular, the role of integrability and its breaking in the dynamical behaviour of a system is of great interest both from a theoretical and an experimental perspective. More generally, the phenomena of chaos which characterises extreme sensitivity to arbitrarily
small perturbations in initial conditions has been extensively studied both in classical and quantum systems. Recently, long-ranged systems have taken a special place as a platform for studying collective behaviour and have become a promising avenue for experimental research.  Notable examples of  long-ranged systems include, one dimensional one component plasma \cite{baxter1993statistical, dhar2018extreme,dhar2017exact,rojas2018universal}, Dyson's Log gas \cite{dyson1962statistical}, Calogero-Moser Systems \cite{calogero1975exactly, calogero1971solution, polychronakos2006physics}, dipolar Bose gas \cite{lu2011strongly, griesmaier2005bose}, ionic systems \cite{brown2003rotational, zhang2017observation,yan2016exploring}, 3D Coulomb gas confined in one dimension \cite{dubin1997minimum}, Yukawa gas \cite{cunden2018universality} to name a few.  Two main ingredients for understanding sensitivity to initial conditions in long-ranged systems are (i) the availability of a family of long-ranged models which contain in them both generic and integrable cases with preferably having reasonably well understood classical and quantum limits and (ii) the availability of diagnostics which can characterize dynamical phenomena and has both classical and quantum counterparts.

The  Riesz gas \cite{riesz1938integrales, agarwal2019harmonically} is one such platform which encompasses a family of long ranged models and we consider the case when it is trapped in an external trapping potential. This family contains in it several models which have themselves been a subject of great interest both from a physics and mathematics perspective. Famous examples for specific values of $k$ include  Dyson's log gas ($k \to 0$), Integrable Calogero-Moser system ($k=2$), 1D One Component Plasma (k=-1), Coloumb gas confined to 1D ($k=1$), dipolar gas ($k=3$) and hard rods ($k\to \infty$).  The parameter $k$ which characterises the power-law interaction spans from ``relatively long-ranged" to ``relatively short-ranged" as we increase $k$. Recently, collective field theory \cite{agarwal2019harmonically} has been provided for the  Riesz gas and its finite ranged generalization \cite{kumar2020particles}. Such a collective description is an important step forward to study nonlinear hydrodynamics \cite{kulkarni2012hydrodynamics,joseph2011observation}. 

The second ingredient, i.e., a suitable diagnostic which can characterize dynamical phenomena is the classical version of the quantum Out-of-Time Ordered Correlator (OTOC) \cite{sekino2008fast,shenker2014black,rozenbaum2017lyapunov,kukuljan2017weak,bohrdt2017scrambling,lakshminarayan2019out,khemani2018velocity} which quantifies growth/decay of perturbations in time and their spread in space. This quantity is precisely suited to explore questions on chaos, aspects of integrability, effect of entropy and nonlinearity to name a few.  In recent years, classical OTOC has been employed as an insightful diagnostic tool to study various extended classical systems such as classical one dimensional spin chains~\cite{Das2018SpinChain}, thermalised fluid obeying truncated  Burgers equation~\cite{Kumar2019},  classical interacting spins
on Kagome lattice \cite{bilitewski2018temperature,bilitewski2020classical}, disordered systems \cite{kumar2020transport}, two-dimensional anisotropic XXZ model \cite{ruidas2020many}, discrete non-linear Schrodinger equation \cite{amit_unpublished} and open systems such as the driven-dissipative duffing chain\cite{Chatterjee2020Duffing}. 

It is worth noting that, to the best of our knowledge, all works on classical OTOC so far have been restricted to the case of short ranged interactions (essentially nearest neighbour) and away from any integrable points. In fact, even for short ranged models, although there have been studies of OTOC in quantum integrable systems \cite{gopalakrishnan2018hydrodynamics, khemani2018velocity}, there has been no work reported on OTOC in classical integrable models to the best of our understanding. In this work, we aim to fill this important gap in our understanding by studying low temperature OTOC of a family of power-law models. Our key results can be summarized as follows: (i) Found exact analytical computation of the dispersion relation in absence of external harmonic potential and utilised it to compute analytical results for OTOC at low temperatures and short enough times (ii) Performed direct numerics and demonstrated excellent agreement with the results obtained after using dispersion relation. (iii) Obtained exact results for OTOC at the integrable point ($k=2$). (iv) Introduced a field theory approach that  paved an alternate path to the investigative aspects of OTOC. 

%
%
%

\textit{Model and definitions:} We consider $N$ classical particles in 1D with pairwise interaction confined by an external harmonic trap. This is the so called Riesz gas \cite{riesz1938integrales} given by,
\begin{subequations}\label{Hamiltonian}
\begin{align}
H =& \sum_{i=1}^{N} \frac{p_{i}^{2}}{2 m} +V_k(\{x_j\})\\
V_k(\{x_j\})=&\sum_{i=1}^{N}\left[\frac{m\omega^2}{2}x^2_i+\frac{J}{2}\sum_{j\neq i}\frac{1}{|x_i-x_j|^k}\right]
\end{align}
\end{subequations}
Here $x_{i}$ is the position of the $i^{th}$ particle (such that $i<j\iff x_{i}<x_{j}$, i.e., ordering is maintained), $p_{i}$ is the corresponding conjugate momentum, $m$ is the mass of each particle, $J$ is the interaction strength and $\omega$ is the frequency of the external trap. Therefore, equations of motion become $\dot{x}_{i} = p_i/m$ and 
\noindent \begin{eqnarray}\label{HamiltonEquations}
\dot{p}_{i}= -m\omega^2 x_{i}+\frac{Jk}{2}\sum_{j\neq i}\frac{ \mathrm{sgn}(x_{i}-x_{j})}{\left|x_{i}-x_{j}\right|^{k+1}} 
\end{eqnarray} 

For a set of initial conditions $\{x_i (0), p_i(0) \}$, one can in principle solve the above $N$ ordinary differential equations each consisting of $(N-1)$ pairing terms thereby rendering it highly nonlocal and nonlinear.\\

\textit{Dispersion relation:} In the ground state in the absence of external trap, we find the dispersion relation (using small oscillation analysis) to be \cite{supp}, 
\begin{equation}\label{particle_DispRel}
\omega_k (q)=\sqrt{\frac{Jk(k+1)}{ma^{k+2}}\big[2\zeta(k+2)-P(k,q)\big]}
\end{equation}
where $P(k,q) = L_{k+2}(e^{-iqa})+L_{k+2}(e^{iqa})$ with $L_{n}(z) = \sum_{p=1}^{\infty} z^p/p^n$ being the Polylogarithm function.
Here $a$ is the lattice spacing/inverse density (i.e., equilibrium is achieved when $x_i(t) = ai$ and $p_i(0)=0$) and $\zeta(z) = \sum_{n=1}^{\infty} 1/n^z$ is the Riemann zeta function. The lattice spacing `$a$' can be introduced in the following manner. Let us say that we have $N$ particles confined in a Harmonic trap of frequency $\omega$. The minimum energy configuration is such that the density takes a \textit{dome} shape\cite{Agarwal2019} and the inter-particle distance at the centre is given by\cite{Agarwal2019}  $a = 2^{\frac{3k+4}{k^2+2 k}}  \big(J(k+1) \zeta (k)/m\big)^{\frac{1}{k+2}} \big(\frac{\omega N}{B\left[1+\frac{1}{k},1+\frac{1}{k}\right]}\big)^{-\frac{2}{k+2}}$ where $B[a,b] = \int_0^1 dw \,w^{a-1}(1-w)^{b-1}$ is the standard Beta function. The homogeneous limit can be thus realised by the careful limit $N\to \infty$ and $\omega \to 0$ keeping $\omega N$ to be a constant. 

It turns out that the above dispersion relation (Eq.~\ref{particle_DispRel}) is periodic with period $2\pi/a$ and it has a maxima at $q=\pi/a$. Using the remarkable property of Polylogarithm function \cite{szego1954erdelyi}, $L_{n}(z)=\Gamma(1-n)\log\big(1/z \big)^{n-1} +\sum_{l=0}^{\infty} \zeta(n-l)\frac{\log(z)^l}{l!}$ for $n \not\in\mathbb{Z}$ and $|\ln z| <2\pi$, we find that the above exact dispersion relation (Eq.~\ref{particle_DispRel}) for $k>1$ has the following small-$q$ expansion~\cite{supp} (up to the next leading relevant order only),

\begin{equation}\label{Omega_expansion}
\begin{aligned}
	\omega_k(q)\approx\begin{cases}
	\alpha_k q-\beta_kq^{k}, \quad\qquad\, \text{ $1<k<3$}\\
	\alpha_k q+\gamma_3q^{3}\log(qa ), \;\quad \text{ $k=3$}\\
	\alpha_k q-\delta_kq^{3}-\beta_kq^{k}, \:\quad \text{ $3<k<5$}\\
	\end{cases}
\end{aligned}
\end{equation}
with 
\begin{equation}
\label{eq:alphabetagammadelta}
\begin{aligned}
\alpha_k=&\sqrt{\frac{Jk(k+1)}{ma^{k}}\zeta(k)},\quad \gamma_3=\frac{1}{4}\sqrt{\frac{Ja}{3m\zeta(3)}}\\ 
\delta_k=&\frac{1}{24}\sqrt{\frac{Jk(k+1)}{m
		\zeta(k)}} \frac{\zeta(k-2)}{a^{(k-4)/2}}\\
\beta_k=&
\sqrt{\frac{Jk(k+1)}{m\zeta(k)}}\cos\left(\frac{\pi}{2}[k+1]\right)a^{k/2-1}\Gamma(-1-k)
\end{aligned}
\end{equation}
where $\Gamma(z)$ is the gamma function. For $k \in \mathbb{Z} $ one can resort to the conventional definition of Polylogarithm function to get the above small-$q$ expansion \cite{supp}. Note also that in the regime $3<k<5$, we wrote down the next-to-next leading order term (Eq.~\ref{Omega_expansion}) since this is what remarkably results in power-law asymptotic behaviour of the OTOC to be discussed later (hence, a relevant term). \\


\begin{figure}[t!]
	\centering
	\begin{subfigure}[t]{\textwidth}
		\centering
		\includegraphics[width=0.45\textwidth, height=7cm, left]{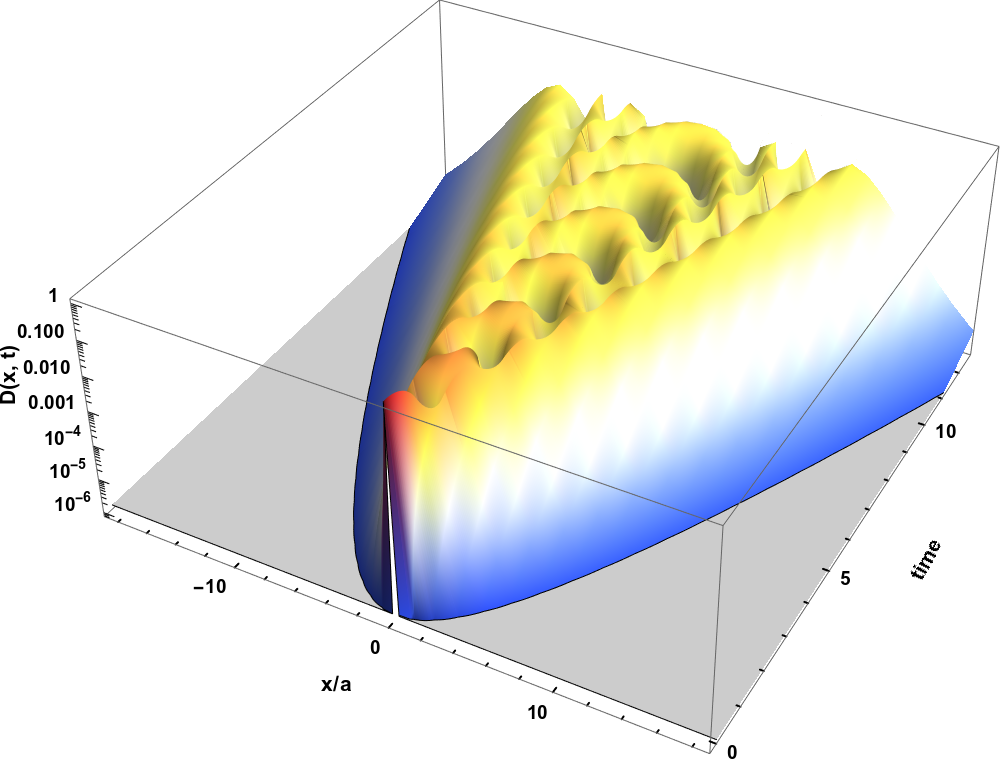}
	\end{subfigure}%
	\\[0.4cm]

	\begin{subfigure}[t]{\textwidth}
		\centering
		\includegraphics[width=0.42\textwidth,height=6cm, left]{N_4097_k_2_PolyLog.png}
	\end{subfigure}
	
	\caption{(Top) Heat map of the OTOC (for clear visualization, we choose $N=65$) from direct numerical simulation  of Eq.~\ref{OTOC} for the Calogero-Moser case ($k=2$). The line traced by the yellow boundary depicts the light cone.  Time axis is in units of $a/v_{B}$. (Bottom) Comparing OTOC from direct numerical simulation of Eq.~\ref{OTOC} with the analytical expression in Eq.~\ref{OTOC_k=2} at a time snapshot, $t\approx 164$ (in units of $a/v_{B}$). Note that, for visualisation purpose,  we have plotted only the positive x-axis and the results are mirror symmetric on the other side. The time $t$ is chosen such that the front moves about $10\%$ from the center to make sure that we are far enough from the edge of the cloud. Here, $a=0.0347$, $v_B=90.5207$.}
	\label{k2fig}
	
\end{figure}
\textit{OTOC:}
The key diagnostic for us is the classical version of the well known quantum OTOC. 
In the Heisenberg picture, the quantum OTOC can be defined \cite{} as $D(x, t)=\langle [\hat{\mathcal{A}}_{x}(t), \hat{\mathcal{B}}_{0}(0)]^{2}\rangle$
where $\hat{\mathcal{A}}_{x}(0)$ and $\hat{\mathcal{B}}_{0}(0)$ are local operators at position $x$ and the origin, respectively. The average $\braket{...}$, is over a given quantum state. This quantity captures the effect of an operator $\hat{\mathcal{B}}_{0}(0)$ on another operator $\hat{\mathcal{A}}_{x}(t)$ at a different position and time. We now replace the commutator by a Poisson bracket, $\{\mathcal{A}_x (t), \mathcal{B}_0(0)\}$. For our purposes, if one makes the identification, $\mathcal{A}_x(t) \equiv x_i (t)$ and $\mathcal{B}_0(0) \equiv p_0(0)$, then the Poisson bracket is  $\{x_{i}(t), p_{0}(0)\}\approx\frac{\delta x_{i}(t)}{\delta x_{0}(0)}$. Therefore, the classical OTOC in our variables becomes $D(i, t)= \langle \{x_i (t), p_0(0)\}^2 \rangle   \approx \langle \big|\frac{\delta x_{i}(t)}{\delta x_{0}(0)}\big|^2 \rangle$. Here, $\braket{...}$ is the average over a thermal ensemble of initial conditions at a given temperature  $T$. However, for low enough temperatures  ($k_B T \ll  J/a^k$ where $k_B$ is the Boltzmann constant), 
the initial conditions are very close to the true ground state (global minima) which is characterised by the set $\{x_i(0)=y_i, p_i(0)= 0\}$ that minimises the energy in Eq.~\ref{Hamiltonian}. Therefore, for low enough temperature we do not need to make an ensemble average. $D(i,t)$ can be interpreted as follows: Take two copies ($I$,$II$) of a system with identical initial conditions. Now, we infinitesimally perturb the position (by $\epsilon$) of one particle (say the middle one) in only one of the copies. In such a case, the OTOC is,
\begin{equation}\label{fullOTOC}
D(i, t) = \bigg|\frac{x_{i}^{I}(t)-x_{i}^{II}(t)}{x_{\frac{N+1}{2}}^{I}(0)-x_{{\frac{N+1}{2}}}^{II}(0)}\bigg|^{2}
=\bigg|\frac{\delta x_i(t)}{\epsilon}\bigg|^{2}
\end{equation}
where we assume $N$ is an odd integer just for convenience. Since, here we are in a regime of sufficiently low temperature, one can invoke a Hessian description, which yields, $\ket{\boldsymbol{\delta \ddot{x} (t)}}=-\boldsymbol{M}\ket{\boldsymbol{\delta x }(t)}$ where $\ket{\boldsymbol{\delta x} (t)}$ is a $N \times 1$ coloumn vector consisting of elements $\delta x_i(t)$ and $\textbf{M}$ is a $N \times N$ Hessian matrix given by $\textbf{M}_{ij}=\left[\frac{\partial^{2}V_k}{\partial x_{i}\partial x_{j}}\right]_{\textbf{x}=\textbf{y}}$. 
In this Hessian limit, Eq.~\ref{fullOTOC} becomes~\cite{supp}  
\begin{equation}
D(i,t)=\left|\sum_{\alpha=1}^{N}\braket{\boldsymbol{\lambda}_{\alpha}\left|\right.\frac{N+1}{2}}\braket{\boldsymbol{e}_{i}\left|\right.\boldsymbol{\lambda}_{\alpha}}\cos\left(\omega_{\alpha}t\right)\right|^{2}\label{OTOC}
\end{equation}
where $\ket{\boldsymbol{\lambda}_{\alpha}}$ is the $\alpha^{th}$ eigenvector of $\textbf{M}$ and $\omega_{\alpha}^2$ is the corresponding eigenvalue. The set ${\{\ket{\boldsymbol{e}_{i}}\}}$ is the standard basis for $\mathbb{R}^{N}$. We have chosen the initial conditions, $\braket{\boldsymbol{e}_{i}|\boldsymbol{\delta x }(t=0)}=\epsilon\delta_{i,\frac{N+1}{2}}$ and  $\braket{\boldsymbol{e}_{i}|\boldsymbol{\delta \dot{x} }(t=0)}=0$  such that $D(i, 0)=\delta_{i,\frac{N+1}{2}}$. In general, neither the equilibrium positions $\{y_i \}$, nor the eigenvectors/eigenvalues are easy to find. Barring the exceptional integrable case \cite{agarwal2019some} of $k=2$, we resort to direct numerics. \\

\begin{figure}[t!]
	\centering
	\begin{subfigure}[t]{\textwidth}
		\centering
		\includegraphics[width=0.45\textwidth, height=7cm, left]{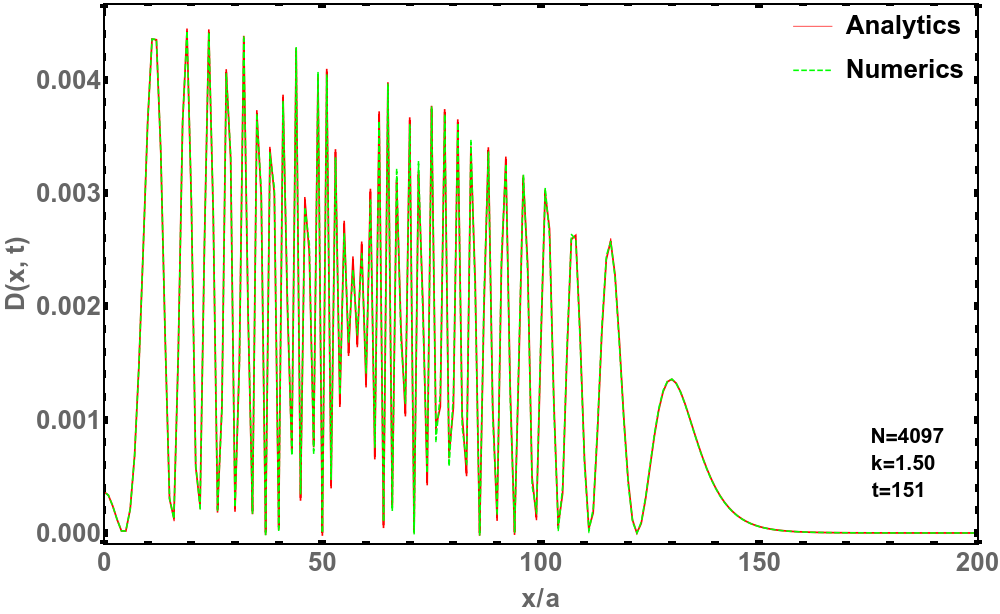}
	\end{subfigure}%
	\\[0.4cm]
	
	\begin{subfigure}[t]{\textwidth}
		\centering
		\includegraphics[width=0.45\textwidth,height=6cm, left]{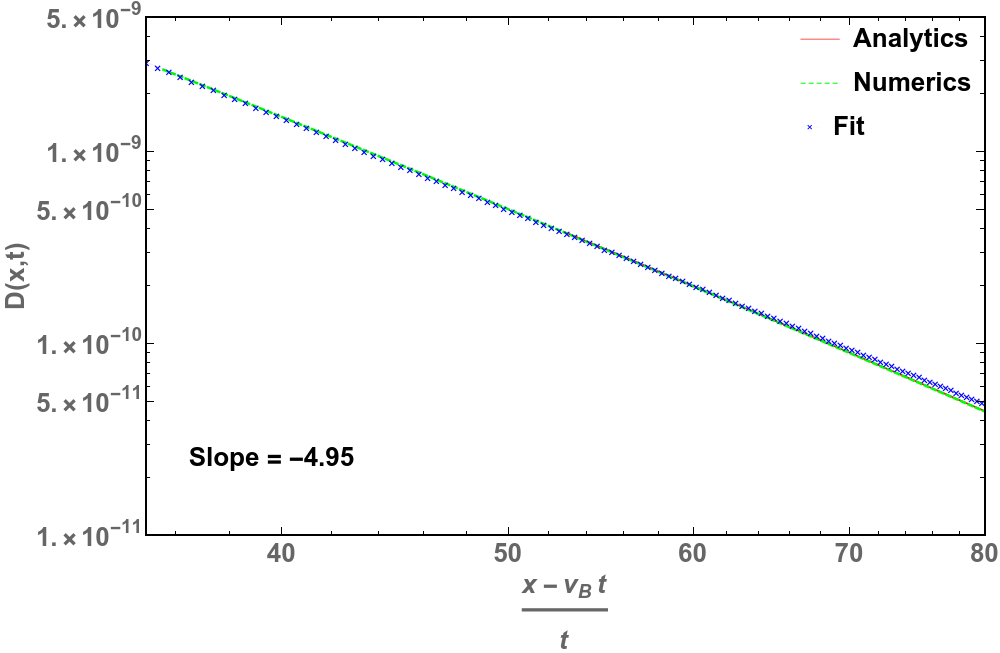}
	\end{subfigure}
	
	\caption{(Top) Comparing OTOC ($k=1.5$) from direct numerical simulation of Eq.~\ref{OTOC} with the analytical expression in Eq.~\ref{OTOC_integral} at a time snapshot, $t\approx 151$ (in units of $a/v_{B}$). Note again that, for visualisation purpose,  we have plotted only the positive x-axis and the results are mirror symmetric on the other side. The time $t$ is again chosen such that the front moves about $10\%$ from the center to make sure that we are far enough from the edge. Note the difference in the profiles of the OTOC. While $k=2$ has a relatively flat envelope (bottom panel of Fig.~\ref{k2fig}), $k=1.5$ shows a downward trending envelope and $k=4.5$ has an upward trending envelope \cite{supp}.  (Bottom) Log-Log plot showing the asymptotic power-law behaviour for $k=1.5$. This plot is the zoomed version of the plot in the top panel near the right front, with $x \in [5.5,\, 8.5]$. Here, $a \sim 0.02237$ and $v_B \sim 53.9296$.}
	\label{fig:k1p5a}
\end{figure}

\textit{Direct numerics:}  In order to compute the OTOC (Eq.~\ref{OTOC}) numerically, we first need to find the set $\{y_i \}$. This is done via the Broyden–Fletcher–Goldfarb–Shanno (BFGS) algorithm\cite{fletcher1970new, broyden1970convergence,supp} which is an efficient way for energy minimization of Eq.~\ref{Hamiltonian} when $N$ is large. All numerical results presented here are for $N=4097$. After getting the positions of the minima ($\{y_i \}$) one can compute the Hessian matrix $\textbf{M}$, its eigenvectors ($\ket{\boldsymbol{\lambda}_{\alpha}}$) and eigenvalues ($\omega_{\alpha}^2$) thereby aiding the computation of OTOC (Eq.~\ref{OTOC}). 
It is also important to mention that for sufficiently large-$N$ the resulting density profile is sufficiently flat near the centre. Hence, for comparing these direct numerical results with analytics (discussed later), we can ignore the harmonic trap as long as we are studying features relatively far from the edges.

The dispersion relation (Eq.~\ref{particle_DispRel}), along with a plane wave ansatz, gives us \cite{Chatterjee2020Duffing,supp}, 
\begin{equation}\label{OTOC_integral}
	D(x,t)=\left|\frac{a}{2\pi}\int_{-\frac{\pi}{a}}^{\frac{\pi}{a} }dq \cos\bigg(qx-\omega_k(q) t\bigg)\right|^{2}
\end{equation}
where $\omega_k(q)$ is given in Eq.~\ref{particle_DispRel}. It is to be noted that the limits of the integral are chosen to be where the dispersion relation reaches a maximum (zero group velocity). Eq.~\ref{OTOC_integral} encodes both the left and the right movers. In other words, Eq.~\ref{OTOC_integral} can be split into two pieces of integral comprising of negative $(-\pi/a,0)$ and positive $(0,\pi/a)$ momentum. 
Next, we will present the results for various values of $k$. Owing to a rich mathematical structure rooted in integrability, we will present the $k=2$ case first. \\


\textit{$k=2$ (Integrable Calogero-Moser Model):} It remarkably turns out that, when $k=2$, the expansion of Eq.~\ref{particle_DispRel} in $q$ terminates to exactly yield, $\omega_{2}(q)=\sqrt{\frac{J}{m}}\left(\frac{\pi q}{a}-\frac{q^{2}}{2}\right)$.
Using this, the OTOC Eq.~\ref{OTOC_integral} gives~\cite{supp}, 

\begin{equation}\label{OTOC_k=2}
	D(x,t)= \left|\frac{a}{2\pi}\left[D_{\mathrm{R}}(x, t)+D_{\mathrm{L}}(x, t)\right]\right|^{2}
\end{equation}
where, $D_{\mathrm{L}}/D_{\mathrm{R}}$ are the left and right moving perturbations respectively and are given by, 

\begin{eqnarray}
\label{eq:dRL}
D_{\mathrm{R,L}}&=&\frac{\sqrt{\pi}}{t u_t}\cos\left(\frac{\eta_{\mp}^{2}}{2u_{t}^{2}}\right)\left[\pm\mathcal{C}\left(\frac{v}{u_{t}\sqrt{\pi}}\right)\mp\mathcal{C}\left(\frac{\eta_{\mp}}{u_t\sqrt{\pi}}\right)\right]\nonumber\\
&+&\frac{\sqrt{\pi}}{t u_t}\sin\left(\frac{\eta_{\mp}^{2}}{2u_{t}^{2}}\right)\left[\pm\mathcal{S}\left(\frac{v}{u_{t}\sqrt{\pi}}
\right)\mp\mathcal{S}\left(\frac{\eta_{\mp}}{u_t\sqrt{\pi}}\right)\right] \nonumber \\
\end{eqnarray}
where $u_{t}=\big(\sqrt{J}/\sqrt{m}t\big)^{1/2}$ and $\mathcal{C}(y)$ and $\mathcal{S}(y)$ are the Fresnel Cosine and Fresnel Sine integrals respectively \cite{supp}. Here $v=x/t$ and $\eta_{\pm}=v \pm \sqrt{J/m}(\pi/a)$. In Fig.~\ref{k2fig}, we show the OTOC for the case $k=2$. Exact agreement between direct numerics (Eq.~\ref{OTOC}) and analytical expression (Eq.~\ref{OTOC_k=2}) is established in Fig.~\ref{k2fig}. The slope of the heat map (top panel in Fig.~\ref{k2fig}) is precisely the butterfly velocity (which in this ground state case is the sound speed) given by $v_B =\pi/a $. This is consistent with the position of the front in the bottom panel of Fig.~\ref{k2fig} which occurs at $v_B \tau$. Extensive asymptotic analysis of Eq.~\ref{OTOC_k=2} and Eq.~\ref{eq:dRL} is presented in the supplementary material~\cite{supp}. For example, we have shown that the right moving perturbation front has the asymptotic behaviour $D_{\mathrm{R}}(t) \sim 1/\eta_-^3$ for large $\eta_->0$.

Although the exact analytical form of the OTOC for $k \neq 2$ is not available due to complexity of the dispersion relation (Eq.~\ref{particle_DispRel}), 
significant advance can be made due to the small-$q$ expansion (Eq.~\ref{Omega_expansion}) which is what we do next. \\

\textit{Non-Integrable case, $k \neq 2$:} 
For the case of $k \neq 2$, one can compare the direct numerical simulations of  Eq.~\ref{OTOC} with the analytical expression in Eq.~\ref{OTOC_integral} with $\omega_k(q)$ given by Eq.~\ref{particle_DispRel}. In Fig.~\ref{fig:k1p5a}, we present an example for $k=1.5$. We see perfect agreement in entire space-time. However, given the complexity of $k\neq 2$ case, we do not have an analog of Eq.~\ref{OTOC_k=2} and Eq.~\ref{eq:dRL} which assumed a fully exact dispersion relation which takes a remarkably simple form in the Calogero-Moser ($k=2$) case.  We therefore resort to a small-$q$ expansion of Eq,~\ref{particle_DispRel}. We recollect that such an expansion gave us Eq.~\ref{Omega_expansion} along with definitions given in Eq.~\ref{eq:alphabetagammadelta}. Note that such an expansion is expected to work close to a particular front (either right or left). Therefore, in the following discussions, we will restrict ourselves to the right sector and our analysis straightforwardly holds for the left sector. We will discuss three cases in Eq.~\ref{Omega_expansion} separately. \\

\textit{$1<k<3 : $} Here, we  simplify Eq.~\ref{OTOC_integral} using the first line of Eq.~\ref{Omega_expansion}. Doing so, we get~\cite{supp}, 
\begin{equation}
D_\mathrm{R}(x,t)=\frac{a}{2\pi} \frac{B_k\left(\Delta_{k, -}\right)}{(3t\beta_k)^{1/k}},\,\,\text{with}\,\Delta_{k, -}=\frac{x-\alpha_k t}{\big(3t\beta_k\big)^{1/k}}
\end{equation}
and the special function is defined as (for $1 < k < 3$), 
\begin{equation}
\label{Bfunc0}
B_k\left(y\right):=\int_{0}^{(3t\beta_k)^{1/k}\frac{\pi}{a}}ds\cos\left(ys+\frac{s^{k}}{3}\right)
\end{equation}
Note that when the lattice spacing $a \to 0$ (which is essentially the large-$N$ limit), then the upper limit of the integral in Eq.~\ref{Bfunc0} becomes $+\infty$. This can be thought of as a $k\neq 3$ generalization of the Airy Integral. In stark contrast to the Airy Integral, in the regime $1<k<3$, we find, $B_k(y) \sim 1/y^{k+1}$ for large $y$. This implies that, $D_{\mathrm{L,R}}(x, t)\propto 1/\Delta_{\pm}^{k+1}$ where $\pm$ indicates whether we are probing the left or the right front respectively. In the bottom panel of Fig.~\ref{fig:k1p5a}, we demonstrate that this powerlaw prediction (for $k=1.5$ which will give $D\sim 1/\Delta_{\pm}^{5}$) is consistent with direct numerical results. \\

\textit{$3<k<5$:}  In this case, if we consider the lowest and the next order term in the dispersion relation (Eq.~\ref{Omega_expansion}) we get, 
	\begin{equation}\label{OTOC_Airylimit}
	D_\mathrm{R}(x, t)=\frac{a \mathrm{Ai}\left(\Delta_{k,-}\right)}{2\big(3t\delta_k \big)^{1/3}},\,\, \text{with } \Delta_{k, -}=\frac{x-\alpha_k t}{\big(3t\delta_k \big)^{1/k}}
	\end{equation}
	where $\mathrm{Ai} (z)$ is the Airy function. This would imply that the asymptotic behaviour of the OTOC (Eq.~\ref{OTOC_integral}) would be characterised by exponential since large argument behaviour of Airy function is $\mathrm{Ai}(z)\sim\frac{e^{-\frac{2}{3}z^{3/2}}}{2\sqrt{\pi}z^{1/4}}$. However, this is because we stopped at $O(q^3)$ in Eq.~\ref{Omega_expansion}. Knowing that we have a power-law model, we expect that the asymptotic behaviour of OTOC should be characterised by power-laws. It turns out that this is captured by considering higher orders in the dispersion relation (Eq.~\ref{OTOC_integral}). For $3<k<5$, the next order term after $q^3$ in Eq.~\ref{Omega_expansion} would be $q^k$ and this will yield a power law tail $D_{\mathrm{L,R}}(x, t)\propto 1/\Delta_{\pm}^{k+1}$. For e.g., for the case $k=4.5$, we demonstrate \cite{supp} that $D_{\mathrm{L,R}}(x, t)\propto 1/\Delta_{\pm}^{5.5}$.
	
We briefly comment on the case $5<k<\infty$ and $k \notin \text{odd integer} $. 	
The term in the dispersion expansion that results in power law is $ \delta_k q^{k}$ which will again yield $D_{\mathrm{L,R}}(x, t)\propto 1/\Delta_{\pm}^{k+1}$.  However, to see this power law, one needs to go to very large asymptotic values since this will happen only after all the exponential behaviours (arising due to $O(q^{2 \mathbb{Z}+1})$ where $\mathbb{Z}$ are positive integers) are suppressed. Next, we will discuss the case when $k$ is an odd integer. In particular, we will discuss the case of $k=3$ but our method can be adapted for all odd integers.
%
%

\textit{$k=3$:}  In this  case, we see a logarithm term in Eq.~\ref{Omega_expansion}, i.e., $q^3 \log(q)$. Note that without the logarithmic piece, we would have ended up with Airy function for the OTOC which would have resulted in exponential asymptotics. However, we now get, 
\begin{equation} 
D_\mathrm{R}(x,t)=\frac{a}{2\pi} \frac{B_3\left(\Delta_{k, -}\right)}{(3t\gamma_3)^{1/3}}
\end{equation}
where, 
\begin{equation}
\label{Bfunc}
B_3\left(y\right):=\int_{0}^{(3t\gamma_3)^{1/3}\frac{\pi}{a}}ds\cos\left(ys+\frac{s^{3}}{3}\log\left[\frac{s}{(3t\gamma_3)^{1/3}}\right] \right)
\end{equation}
We find that the large $y$ behaviour is $B_3\left(y\right) \sim 1/y^4$ which implies  $D_{\mathrm{L,R}}(x, t)\propto 1/\Delta_{\pm}^{4}$. It is remarkable to note that the expected power-law behaviour is recovered as a result of the intricate role played by the Logarithmic term in the dispersion relation for $k=3$.  We also find that for odd integers, i.e., $k \in 2 \mathbb{Z}+1$ where $\mathbb{Z}$ are positive integers, the power-law is recovered by a term in the dispersion relation of the form $\beta_k q^k \log(q)$. \\

\textit{Field theory:} An alternative approach to studying the large $N$ behaviour of this system is to investigate the collective field theory. Recently, a systematic derivation of large-$N$ field theory  \cite{Agarwal2019} was achieved. Here, we will show that certain aspects of spatio-temporal spread of correlations such as the butterfly velocity can be obtained by a field theory. 
Let us define a density field, $\rho(x) =  \sum_{i=1}^{N} \langle \delta(x-x_i) \rangle$ where $\langle ...\rangle$ denotes an average with respect to a Boltzman measure. We also define a momentum field, $j(x) = \sum_{i=1}^N \langle p_i \delta(x-x_i) \rangle$. We will introduce a velocity field $v(x)$ such that $j(x) = \rho(x) v(x)$. 
In large-$N$ at sufficiently low temperatures, the field theory is given by, 
$H[\rho_N (x)] \approx  \frac{m}{2} \int \rho(x) v(x)^2 dx+ J\zeta(k)  \int  \rho(x)^{k+1} dx$. 
This in conjugation with Poisson Brackets,  $ \{\rho(x_1),v(x_2) \} =  \frac{1}{m} \delta'(x_1-x_2) $ gives us
\begin{equation}
	\dot{\rho}=-\partial_x(\rho v),\,\,
	\dot{v} = -\partial_x \bigg[\frac{v^2}{2} +   \frac{J}{m}\zeta(k) (k+1) \rho^k  + .... \bigg]
\end{equation}	
One can linearise the above Continuity and Euler equations by 
using $\rho(x, t)=\rho_{0}  + \delta \rho (y,t)$, $v (x,t) = 0+ \delta v (x,t)$
to get a wave equation with sound speed given by $c= \sqrt{(J/m) k(k+1) \zeta(k) \rho_0^k }$. This is precisely the butterfly velocity in agreement with $\alpha_k$ given in Eq.~\ref{eq:alphabetagammadelta}. The background density ($\rho_0$) or the inverse lattice spacing ($a^{-1}$) has already been discussed before.
\textit{Conclusions:} In summary, we studied a family of power-law models at low temperature.  In particular, we probed in detail the spatio-temporal spread of perturbations.  We could analytically compute the dispersion relation in absence of external harmonic potential and utilise it to get  analytical results for OTOC at low temperatures. We then performed direct numerics with the aid of BFGS algorithm and demonstrated excellent agreement with the results obtained after using dispersion relation. Exact results for OTOC at the integrable point ($k=2$) were obtained. We also presented a collective field theory approach to understand certain features of the OTOC such as the butterfly speed. 

Our work was restricted to low enough temperatures and short enough times so that the system is still in the linear regime.  Therefore, although there was spread of perturbations, we restricted ourselves to temperatures and time scales where there was no growth in magnitude of perturbations. Precisely quantifying the limits of this regime as well as exploring beyond it is part of our planned future work. 
Being a long-ranged system of particles, high temperature studies are considerably numerically intense.  In such high temperature cases one expects exponential (non-integrable, $k\neq 2$) or power-law (integrable, $k=2$) growth and this will be addressed in a future work. Understanding aspects of integrability ($k=2$) and its breaking through the lens of OTOC still remains largely unexplored and is an interesting future direction. \\

\textit{Acknowledgements:}
We would like to thank A. Dhar, A. Kundu and A. K. Chatterjee for useful discussions. MK would like to acknowledge support from the project 6004-1 of the Indo-French Centre for the Promotion of Advanced Research (IFCPAR), Ramanujan Fellowship (SB/S2/RJN-114/2016), SERB Early Career Research Award (ECR/2018/002085) and SERB Matrics Grant (MTR/2019/001101) from the Science
and Engineering Research Board (SERB), Department of Science and Technology, Government of India.  DH is supported in part by (USA) DOE grant DE-SC0016244.
\bibliography{references.bib}
\end{document}


\title{Supplementary Material for ``Spatio-temporal spread of perturbations in power-law models at low temperatures:
Exact results for OTOC"}
\author{Bhanu Kiran S.}
\address{International Centre for Theoretical Sciences, Tata Institute of Fundamental Research, Bengaluru -- 560089, India}
\author{ David A. Huse}
\address{Department of Physics, Princeton University, Princeton, NJ 08544, USA}
\author{Manas Kulkarni}
\address{International Centre for Theoretical Sciences, Tata Institute of Fundamental Research, Bengaluru -- 560089, India}
\date{\today}
\maketitle

\tableofcontents

\hfill \break
\hfill \break
The contents of this supplementary are as follows: In Sec.~\ref{OTOC hessianbasis} we provide the derivation of the OTOC in the Hessian approximation, valid in the low-$T$ regime.  In Sec.~\ref{derivation disprel} we derive the exact dispersion relation for the Riesz gas family of models in the low temperature and large-$N$ regime. In Sec.~\ref{OTOC planewavebasis} we obtain an expression for the OTOC in the plane wave basis, valid in the large-$N$ regime.
Sec.~\ref{Asymptotics} contains detailed discussion about the asymptotic behaviour of the OTOC outside the light-cone in various regimes of $k$. Finally, in Sec.~\ref{BFGS} we briefly explain the Broyden–Fletcher–Goldfarb–Shanno (BFGS) algorithm which was employed to compute the ground state configurations of this family of models for large number of particles.

\section{OTOC in Hessian approximation valid in the low $T$ regime} \label{OTOC hessianbasis}
Here, we present a derivation of the OTOC in the Hessian approximation. Let us recap that the Hamiltonian is given by
\begin{subequations}\label{Hamiltonian}
\begin{align}
H =& \sum_{i=1}^{N} \frac{p_{i}^{2}}{2 m} +V_k(\{x_j\})\\
V_k(\{x_j\})=&\sum_{i=1}^{N}\left[\frac{m\omega^2}{2}x^2_i+\frac{J}{2}\sum_{j\neq i}\frac{1}{|x_i-x_j|^k}\right]
\end{align}
\end{subequations}
which therefore yields the $N \times N$ Hessian matrix, 
\begin{equation}
\label{eq:hessian}
\textbf{M}_{ij}=\left[\frac{\partial^{2}V_k}{\partial x_{i}\partial x_{j}}\right]_{\textbf{x}=\textbf{y}}
\end{equation}
where \textbf{y} is the equilibrium solution that minimizes Eq.~\ref{Hamiltonian}.  Let us define a coloumn vector ($N \times 1$) representing the relative displacement (again a $N \times 1$ column vector) of corresponding particles of copy $I$ and copy $II$
\begin{equation}
	\ket{\boldsymbol{\delta x }(t)}=\ket{\boldsymbol{x}^{I}(t)}-\ket{\boldsymbol{x}^{II}(t)}
\end{equation}
Let $\boldsymbol{M}$ be the Hessian matrix. We therefore have
\begin{equation}
	\ket{\boldsymbol{\delta \ddot{x} }(t)}=-\boldsymbol{M}\ket{\boldsymbol{\delta x }(t)}
\end{equation}

We assume initial condition for velocities $\braket{\boldsymbol{e}_{i}|\boldsymbol{\delta \dot{x} }(t=0)}=0$ where ${\{\ket{\boldsymbol{e}_{i}}\}}$ is the standard basis for $\mathbb{R}^{N}$. Let $\ket{\boldsymbol{\lambda}_{\alpha}}$ be the eigenvectors and $\omega_{\alpha}^{2}$ be the corresponding eigenvalues of $\boldsymbol{M}$. In the eigenbasis, the solution is 
\begin{equation}
\ket{\boldsymbol{\delta x }(t)}=\sum_{\alpha=1}^{N}\braket{\boldsymbol{\lambda}_{\alpha}|\boldsymbol{\delta x }(0)}\ket{\boldsymbol{\lambda}_{\alpha}}\cos\left(\omega_{\alpha} t\right)
\end{equation}
With our chosen initial condition $\braket{\boldsymbol{e}_{i}|\boldsymbol{\delta x }(0)}=\epsilon\delta_{i,\frac{N+1}{2}}$ the OTOC finally becomes
\begin{equation}
\begin{aligned}
	D(i, t)=&\left|\frac{\braket{\boldsymbol{e}_{i}|\boldsymbol{\delta x}(t)}^{2}}{\epsilon}\right|=\left|\sum_{\alpha=1}^{N}\braket{\boldsymbol{\lambda}_{\alpha}|\frac{N+1}{2}}\braket{\boldsymbol{e}_{i}|\boldsymbol{\lambda}_{\alpha}}\cos\left(\omega_{\alpha} t\right)\right|^{2}
\end{aligned}
\end{equation}

\section{Derivation of the dispersion relation} \label{derivation disprel}
Here, we present a detailed derivation of the dispersion relation. Let us consider the Riesz gas (Eq.~\ref{Hamiltonian}) without the Harmonic trap. We will also consider the case of $N\to \infty$ and label the particles from $-\infty$ to $+\infty$ without loss of generality. 
The corresponding equation of motion are
\begin{equation}
\dot{x}_{i} = p_i/m,\ \quad	\dot{p}_{i}= \frac{Jk}{2}\sum_{\substack{j= -\infty \\ j\neq i}}^{\infty}\frac{ \mathrm{sgn}(x_{i}-x_{j})}{\left|x_{i}-x_{j}\right|^{k+1}} \label{Accl_in_natural_units_notrap}
\end{equation}
The equilibrium solution takes the form 
\begin{equation}\label{equi_soln}
\begin{aligned}
	x_{i}(t=0) =& \,ia, \quad  p_{i}(t=0) =0 
\end{aligned}
\end{equation}
where $a$ is a chosen lattice spacing. The above solution (Eq.~\ref{equi_soln}) does not evolve and therefore it is an equilibrium solution for Eq.~\ref{Accl_in_natural_units_notrap}. 
This is so because Eq.~\ref{Accl_in_natural_units_notrap} is an odd sum. In other words
\begin{equation}\label{odd_infty_sum}
	\sum_{\substack{j=-\infty\\j\ne i}}^{\infty} \frac{\mathrm{sgn}(x_{i}-x_{j})}{\left|x_{i}-x_{j}\right|^{k+1}}=\frac{1}{a^{k+1}}\sum_{\substack{j=-\infty\\j\ne i}}^{\infty}\frac{\mathrm{sgn}(i-j)}{\left|i-j\right|^{k+1}}=0
\end{equation}
Having found an equilibrium background, we do a small oscillation analysis of Eq.~\ref{Accl_in_natural_units_notrap}. Using the ansatz, 
\begin{equation}
\label{eq:small_osc}
x_{i}(t)=ai+\epsilon\cos\left(qai - \omega_k t\right)
\end{equation}
we get from Eq.~\ref{Accl_in_natural_units_notrap} (for convenience of notation we suppress the subscript $k$ in $\omega_k$ in the below calculation and it is restored in the end)
\begin{equation}
\label{eqd1}
-m\omega^{2}\epsilon\cos\left(qai-\omega t\right)= \frac{Jk}{2}\sum_{j\neq i}\frac{\mathrm{sgn}(x_{i}-x{j})}{\big|a(i-j)+\epsilon\left[\cos\left(qai-\omega t\right)-\cos\left(qaj-\omega t\right)\right]\big|^{k+1}}
\end{equation}
Splitting the summation above (Eq.~\ref{eqd1}) gives
\begin{align}\label{broken_sum_ignore_abs}
-m\omega^{2}\epsilon\cos\left(qai-\omega t\right)=&\sum_{j<i}\frac{Jk}{\big|a(i-j)+\epsilon\left[\cos\left(qai-\omega t\right)-\cos\left(qaj - \omega t\right)\right]\big|^{k+1}}\nonumber\\
&-\sum_{j>i}\frac{Jk}{\big|a(i-j)+\epsilon\left[\cos\left(qai-\omega t\right)-\cos\left(qaj - \omega t\right)\right]\big|^{k+1}}\nonumber \\ 
=&\sum_{j<i}\frac{Jk}{\left|a(i-j)\right|^{k+1}}\frac{1}{\big|1+\frac{\epsilon}{a(i-j)}\left[\cos\left(qai-\omega t\right)-\cos\left(qaj - \omega t\right)\right]\big|^{k+1}}\nonumber \\
-&\sum_{j>i}\frac{Jk}{\left|a(i-j)\right|^{k+1}}
\frac{1}{\left|1+\frac{\epsilon}{a(i-j)}\left[\cos\left(qai-\omega t\right)-\cos\left(qaj - \omega t\right)\right]\right|^{k+1}}
\end{align}
Since $\frac{\epsilon}{a}<<1$ (small oscillation theory), we use the Binomial expansion. Remember that our choice of labelling, $x_{i}<x_{j}\iff i<j,$ means that the absolute value can be removed once the sum is broken into parts as done in Eq.~\ref{broken_sum_ignore_abs}. This gives

\begin{align}
-m\omega^{2}\epsilon\cos\left(qai-\omega t\right)=&\sum_{j<i}\frac{Jk}{\left[a(i-j)\right]^{k+1}}
\left(1+\frac{\epsilon}{a(i-j)}\left[\cos\left(qai-\omega t\right)-\cos\left(qaj - \omega t\right)\right]\right)^{-(k+1)}\nonumber \\
&-\sum_{j>i}\frac{Jk}{\left[a(j-i)\right]^{k+1}}
\left(1+\frac{\epsilon}{a(j-i)}\left[\cos\left(qaj-\omega t\right)-\cos\left(qai - \omega t\right)\right]\right)^{-(k+1)}
\end{align}
which keeping only leading order in $\epsilon$ yields
\begin{align}
-m\omega^{2}\epsilon\cos\left(qai-\omega t\right)=&\sum_{j=-\infty}^{i-1}\frac{Jk}{\left[a(i-j)\right]^{k+1}}
\left(1-\frac{\epsilon(k+1)}{a(i-j)}\left[\cos\left(qai-\omega t\right)-\cos\left(qaj - \omega t\right)\right]\right)\nonumber \\
&-\sum_{j=i+1}^{\infty}\frac{Jk}{\left[a(j-i)\right]^{k+1}}
\left(1-\frac{\epsilon(k+1)}{a(j-i)}\left[\cos\left(qaj-\omega t\right)-\cos\left( qai - \omega t\right)\right]\right)
\end{align}
Notice that the first term in each of the sums cancel one another. We are therefore left with 
\begin{align}
m\omega^{2}\cos\left(qai-\omega t\right)=&
\sum_{j=-\infty}^{i-1}\frac{ Jk(k+1)}{\left[a(i-j)\right]^{k+2}}
\left[\cos\left(qai-\omega t\right)-\cos\left(qaj - \omega t\right)\right]\nonumber \\
&-\sum_{j=i+1}^{\infty}\frac{ Jk(k+1)}{\left[a(j-i)\right]^{k+2}}
\left[\cos\left(qaj-\omega t\right)-\cos\left( qai - \omega t\right)\right]
\end{align}
By using a simple trigonometric identity we get
\\
\begin{align}
\label{eq:trig}
m\omega^{2}\cos\left(qai-\omega t\right)=&
\sum_{j=-\infty}^{i-1}\frac{-2Jk(k+1)}{\left[a(i-j)\right]^{k+2}}
\left[\sin\left(qa\frac{i+j}{2}-\omega t\right)\sin\left(qa\frac{i-j}{2}\right)\right]\nonumber \\
&+\sum_{j=i+1}^{\infty}\frac{2Jk(k+1)}{\left[a(j-i)\right]^{k+2}}
\left[\sin\left(qa\frac{j+i}{2}-\omega t\right)\sin\left(qa\frac{j-i}{2}\right)\right]
\end{align}
Let us define integers $i-j \equiv x_{1}$ and $j-i \equiv x_{2}$ in the first and the second sum in Eq.~\ref{eq:trig} respectively. We then get 
\begin{align}
\label{eq:trig1}
m\omega^{2}\cos\left(qai-\omega t\right)=&
\sum_{x_{1}=1}^{\infty}\frac{-2Jk(k+1)}{\left(ax_{1}\right)^{k+2}}\left[\sin\left(\frac{qa}{2}(2i-x_{1})-\omega t\right)\sin\left(\frac{qa}{2}x_{1}\right)\right]\nonumber \\
&+\sum_{x_{2}=1}^{\infty}\frac{2Jk(k+1)}{\left(ax_{2}\right)^{k+2}}\left[\sin\left(\frac{qa}{2}(2i+x_{2})-\omega t\right)\sin\left(\frac{qa}{2}x_{2}\right)\right]
\end{align}
Further simplification of Eq.~\ref{eq:trig1} yields
\begin{align}
\label{eq:trig3}
m\omega^{2}\cos\left(qai-\omega t\right)=&
\sum_{x_{1}=1}^{\infty}\frac{-2Jk(k+1)}{\left(ax_{1}\right)^{k+2}}\left[\sin\left(qai-\omega t\right)\cos\left(\frac{qax_{1}}{2}\right)\sin\left(\frac{qa}{2}x_{1}\right)\right.
\left.-\cos\left(qai-\omega t\right)\sin^{2}\left(\frac{qa}{2}x_{1}\right)\right]\nonumber \\
&+\sum_{x_{2}=1}^{\infty}\frac{2Jk(k+1)}{\left(ax_{2}\right)^{k+2}}\left[\sin\left(qai-\omega t\right)\cos\left(\frac{qa}{2}x_{2}\right)\sin\left(\frac{qa}{2}x_{2}\right)\right.
\left.+\cos\left(qai-\omega t\right)\sin^{2}\left(\frac{qa}{2}x_{2}\right)\right]
\end{align}
In above Eq.~\ref{eq:trig3}, $x_1$ and $x_2$ are dummy variables and it is easy to see that the first terms in each of the sums cancels each other. The expression (Eq.~\ref{eq:trig3}) easily simplifies further to (restoring the subscript $k$ notation from Eq.~\ref{eq:small_osc})
\begin{equation}\label{Disp_rel}
\omega_k^{2}(q)=\frac{4Jk(k+1)}{ma^{k+2}}\sum_{x_{1}=1}^{\infty}\frac{\sin^{2}\left(\frac{qa}{2}x_{1}\right)}{x_{1}^{k+2}}
\end{equation}
Using the relation $\sin\left(z\right)=\frac{e^{iz}-e^{-iz}}{2i}$, 
we see that Eq.~\ref{Disp_rel} can be simplified to 
\begin{equation}
	\omega_k(q)=\sqrt{\frac{Jk(k+1)}{ma^{k+2}}\left[2\zeta(k+2)-L_{k+2}(e^{-iqa})\right.
	\left.-L_{k+2}(e^{iqa})\right]}\label{PolyLog_DispRel}
\end{equation}
where  $\zeta(z) = \sum_{n=1}^{\infty} 1/n^z$ is the Riemann zeta function and $\mathrm{Li}_{n}(z)=\sum_{k=1}^{\infty}\frac{z^{k}}{k^{n}}$ is the Polylogarithm function.  Note that Eq.~\ref{PolyLog_DispRel} is the exact dispersion relation. Fig.~\ref{k2fig} (Left) shows a plot of $\omega_k(q)$ for $k=2$.
Further, using the following remarkable identity of Polylogarithm function \cite{szego1954erdelyi}
\begin{equation}
L_{n}(z)=\Gamma(1-n)\log\big(1/z \big)^{n-1} +\sum_{l=0}^{\infty} \zeta(n-l)\frac{\log(z)^l}{l!}
\end{equation}
for $n \not\in\mathbb{Z}$ and $|\ln z| <2\pi$, one can expand Eq.~\ref{PolyLog_DispRel} (to the next-to-next leading order) as a power series in $q$ to obtain
\begin{equation}\label{Omega_expansion}
\begin{aligned}
\omega_k(q)\approx\begin{cases}
\alpha_k q-\beta_kq^{k}-\delta_kq^{3}, \qquad \qquad\, \text{ $1<k<3$}\\
\alpha_k q+\gamma_3q^{3}\log(q a)- \bar{\gamma}_3 q^{3}, \;\quad \text{ $k=3$}\\
\alpha_k q-\delta_kq^{3}-\beta_kq^{k}, \qquad \qquad \text{ $3<k<5$}\\
\end{cases}
\end{aligned}
\end{equation}
with 
\begin{equation}
\label{eq:alphabetagammadelta}
\begin{aligned}
\alpha_k=&\sqrt{\frac{Jk(k+1)}{ma^{k}}\zeta(k)},\qquad \gamma_3=\frac{1}{4}\sqrt{\frac{Ja}{3m\zeta(3)}}, \qquad 
\delta_k=\frac{1}{24}\sqrt{\frac{Jk(k+1)}{m
		\zeta(k)}} \frac{\zeta(k-2)}{a^{(k-4)/2}}\\
\beta_k=&
\sqrt{\frac{Jk(k+1)}{m\zeta(k)}}\cos\left(\frac{\pi}{2}[k+1]\right)a^{k/2-1}\Gamma(-1-k), \qquad \bar{\gamma_3}= \frac{25\gamma_3}{12} 
\end{aligned}
\end{equation}
It is to be noted that, for $k \in \mathbb{Z} $ one can resort to the conventional definition of Polylogarithm function to get the above small-$q$ expansion.
In Eq.~\ref{eq:alphabetagammadelta}, we see that both $\beta_k$ and $\delta_k$ terms diverge in the limit $k\to3$. But upon a careful computation of this limit, it is seen that these divergences in fact cancel each other and give rise to the logarithm term in the dispersion relation for $k=3$. More precisely, 
\begin{eqnarray}\label{k to 3 limit}
\lim_{k \to 3}  (\beta_kq^{k}+\delta_kq^{3}) &=&  \lim_{\epsilon \to 0}  (\beta_{3 \pm \eta}q^{3 \pm \eta}+\delta_{3\pm \eta}q^{3}) \nonumber \\ &=& \sqrt{\frac{Ja3(4)}{m\zeta(3)}}q^{3}\lim_{\epsilon \to 0}\left(q^{\epsilon}a^{\epsilon/2}\Gamma(-4\mp \epsilon)+\frac{1}{24}a^{-\epsilon/2}\zeta(1 \pm \epsilon)\right)\nonumber \\ &=&6\sqrt{\frac{Ja}{3m\zeta(3)}}\left(\frac{25-12\log(qa)}{288}\right)q^{3}\nonumber\\
&=&\frac{1}{4}\sqrt{\frac{Ja}{3m\zeta(3)}}\left(\frac{25}{12}-\log(a)-\log(q)\right)q^{3} \nonumber \\
&=& -\gamma_3q^{3}\log(q a) + \bar{\gamma}_3 q^{3}
\end{eqnarray}
Moreover, this kind of cancellation is seen for all odd integer values of $k$, leading to a logarithm term $\sim q^k\log(q)$. This is a crucial term in the dispersion relation because this is precisely the term which leads to a power-law decay of the OTOC beyond the cone, for odd integer values of $k$.\\

\section{OTOC in planewave basis for large-$N$} \label{OTOC planewavebasis}
Here, we briefly derive the large-$N$ expression of the OTOC in the plane wave basis. Let, $\ket{\delta\textbf{x}(t)}$ be a vector containing the displacements $\delta x_j(t)$. We can expand this vector in terms of the plane-wave phonons of our system as
\begin{equation}
\label{eq:planewave}
\ket{\delta\textbf{x}(t)}=\frac{a}{2\pi}\int_{-\pi/a}^{\pi/a}dq\,\langle q |\delta\textbf{x}(t)\rangle\ket{q}\implies\langle \boldsymbol{e}_{j} |\delta\textbf{x}(t)\rangle=\frac{a}{2\pi}\int_{-\pi/a}^{\pi/a}dq\,\langle q |\delta\textbf{x}(t)\rangle\langle \boldsymbol{e}_{j}\ket{q}
\end{equation}
where $\langle \boldsymbol{e}_{j}\ket{\delta\textbf{x}(t)}=\delta x_j(t)$. Here we remind that  ${\{\ket{\boldsymbol{e}_{i}}\}}$ is the standard basis for $\mathbb{R}^{N}$. 
At At $t=0$,  since  $\langle \boldsymbol{e}_{j} | \delta\textbf{x}(t=0)\rangle= \epsilon\, \delta_{j, 0}$ and $\langle \boldsymbol{e}_{j}\ket{q}=\text{Real} [e^{i(qja-\omega_k(q)t)}] $ at time $t$, we have  $\langle q |\delta\textbf{x}(t=0)\rangle = \epsilon$ (constant) which ensures that Eq.~\ref{eq:planewave} is satisfied at $t=0$.
The noninteracting phonons simply evolve freely, so,
using $j a \equiv x$, we can write
\begin{equation}
\langle \boldsymbol{e}_{j} | \delta\textbf{x}(t)\rangle=  \frac{\epsilon a}{2\pi} \text{Real} \Big[\int_{-\pi/a}^{\pi/a}dq\, e^{i[qx-\omega_k(q)t]} \Big]
\end{equation}
We therefore finally get, 
\begin{equation}\label{OTOC_integral}
	D(x,t)= \left|\frac{\braket{\boldsymbol{e}_{j}|\boldsymbol{\delta x}(t)}^{2}}{\epsilon}\right| = \left|\frac{a}{2\pi}\int_{-\frac{\pi}{a}}^{\frac{\pi}{a} }dq \cos\bigg(qx-\omega_k(q) t\bigg)\right|^{2}
\end{equation}
where $\omega_k(q)$ is given in Eq.~\ref{PolyLog_DispRel}. Figs.~\ref{k2fig} (Middle) and \ref{k4p5fig} (Middle) show a plot of Eq.~\ref{OTOC_integral} for cases $k=2$ and $k=4.5$ respectively. The figures only show the right sector ($x>0$), the left sector ($x<0$) being just the mirror image.

\section{Asymptotic analysis of OTOC} \label{Asymptotics}

In this section we provide the asymptotic analysis of the OTOC at either the left or the right front. We divide this section into $k=2$ (integrable case) and $k \neq 2$ (non-integrable case). 

\subsection{Integrable case, $k=2$}
\begin{figure}[t!]
		\centering
		\includegraphics[width=0.32\textwidth,height=4.5cm]{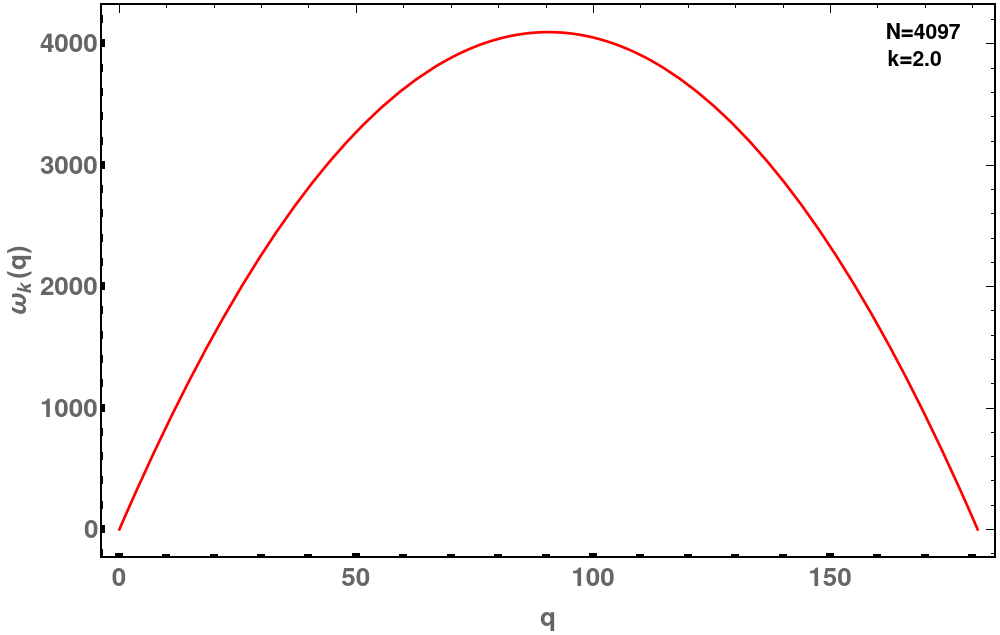}
		\includegraphics[width=0.32\textwidth,height=4.5cm]{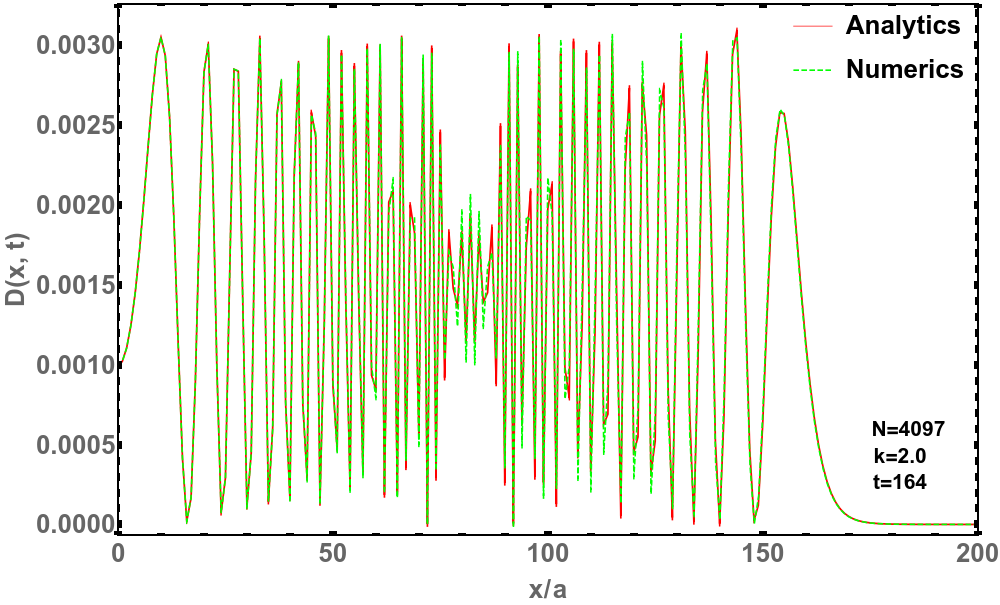}
		\includegraphics[width=0.32\textwidth, height=4.5cm]{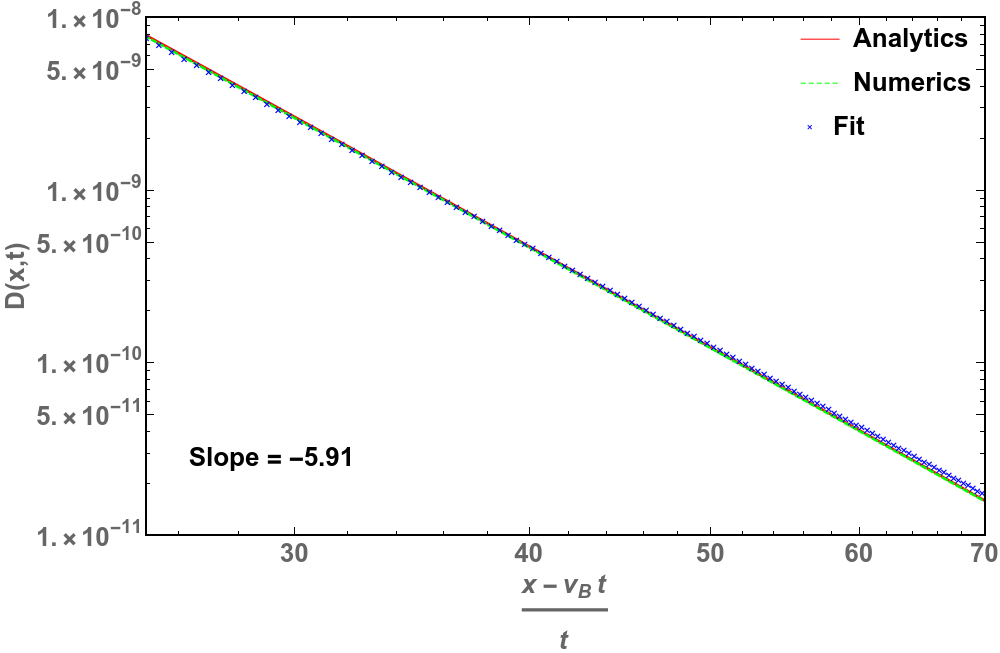}
	
	\caption{  (Left) Dispersion relation over one time period in $q$, for $k=2$, $q\in[0, 2\pi/a]$.   (Middle) Comparing OTOC from direct numerical simulation with the analytical expression at a time snapshot, $t\approx 164$ (in units of $a/v_{B}$). Note that, for visualisation purpose,  we have plotted only the positive x-axis and the results are mirror symmetric on the other side. (Right) Power law decay of OTOC beyond the cone for $k=2$ and $N=4097$. The slope depicts the exponent of the power law. Note that the asymptotics agrees with Eq.~\ref{OTOC_k=2_asymp} including coefficients. Here, $a\sim0.0347$, $v_B\sim90.5207$.}
	\label{k2fig}
	
\end{figure}

In this case, the dispersion (Eq.\ref{PolyLog_DispRel}) remarkably has a finite expansion, 
\begin{equation}
\omega_2(q)=\sqrt{\frac{J}{m}}\left(\frac{\pi q}{a}-\frac{q^{2}}{2}\right),\quad \text{for $k=2$}
\end{equation}
This allows us to cast the OTOC in terms of the Fresnel functions without resorting to any approximations. We would like to understand the behaviour of the OTOC in the large $\eta_{\pm}$ limit where $\eta_{\pm}=v\pm \sqrt{\frac{J}{m}}\frac{\pi}{a}$. For e.g., large $\eta_-$ means that we are probing the asymptotics of the right front.  For this, we look at the asymptotics of the following equations.
\begin{eqnarray}
\label{eq:dRL}
D_{\mathrm{R,L}}&=&\sqrt{\frac{\sqrt{m}\pi}{\sqrt{J}t}}\cos\left(\frac{\sqrt{m}\eta_{\mp}^{2}t}{2\sqrt{J}}\right)\left[\pm\mathcal{C}\left(v\sqrt{\frac{\sqrt{m}t}{\sqrt{J}\pi}}\right)\mp\mathcal{C}\left(\eta_{\mp}\sqrt{\frac{\sqrt{m}t}{\sqrt{J}\pi}}\right)\right]\nonumber\\
&+&\sqrt{\frac{\sqrt{m}\pi}{\sqrt{J}t}}\sin\left(\frac{\sqrt{m}\eta_{\mp}^{2}t}{2\sqrt{J}}\right)\left[\pm\mathcal{S}\left(v\sqrt{\frac{\sqrt{m}t}{\sqrt{J}\pi}}\right)\mp\mathcal{S}\left(\eta_{\mp}\sqrt{\frac{\sqrt{m}t}{\sqrt{J}\pi}}\right)\right]
\end{eqnarray}
where $\mathcal{C}(r) = \int_{0}^{r}dt\cos(\pi t^{2}/2) $ and $\mathcal{S}(r) = \int_{0}^{r}dt\sin(\pi t^{2}/2)$ are the Fresnel Cosine and Sine integrals, respectively.\\
For our purpose we use the large argument asymptotic form of Fresnel integrals. In the large $r$ limit, we have
\begin{subequations}\label{Fresnel_asymptotic_form}
\begin{align}
 	\mathcal{S}(r)=&\frac{1}{2}+\cos\left(\frac{\pi r^{2}}{2}+O(r^{-4})\right)\left[-\frac{1}{\pi r}+O(r^{-4})\right]
 	+\sin\left(\frac{\pi r^{2}}{2}+O(r^{-4})\right)\left[-\frac{1}{\pi^{2}r^{3}}+O(r^{-4})\right]\\
 	\mathcal{C}(r)=&\frac{1}{2}+\sin\left(\frac{\pi r^{2}}{2}+O(r^{-4})\right)\left[\frac{1}{\pi r}+O(r^{-4})\right]
 	+\cos\left(\frac{\pi r^{2}}{2}+O(r^{-4})\right)\left[-\frac{1}{\pi^{2}r^{3}}+O(r^{-4})\right]
\end{align}
\end{subequations}
We re-emphasize here that to get the correct asymptotic behaviour in the right sector, we go to the large $\eta_{-}$ limit. Note that $v\gg\eta_{-}$. This is because $v=v_B + \eta_{-}$ and $v_B$ is very large in the limit that lattice spacing is very small. Therefore, we can take the limit $v\to \infty$ while assuming $\eta_{-}$ to be large enough to be able to do asymptotics. We use Eqs.~\ref{Fresnel_asymptotic_form} to get the large $\eta_{-}$ expansion for $D_R$ and set $v\sqrt{\frac{mt}{J\pi}}\to \infty$ in the argument of the Fresnel functions. This finally yields
	\begin{align}\label{Fresnel_asymp}
		D_{\mathrm{R}}\approx&\sqrt{\frac{2\sqrt{m}}{\sqrt{J}t}}\cos\left(\frac{\sqrt{m}\eta_{-}^{2}t}{2\sqrt{J}}\right)
		\sqrt{\frac{\pi}{8}}
		+\sqrt{\frac{2\sqrt{m}}{\sqrt{J}t}}\sin\left(\frac{\sqrt{m}\eta_{-}^{2}t}{2\sqrt{J}}\right)
		\sqrt{\frac{\pi}{8}}\nonumber\\
		&-\sqrt{\frac{2\sqrt{m}}{\sqrt{J}t}}\cos\left(\frac{\sqrt{m}\eta_{-}^{2}t}{2\sqrt{J}}\right)\left[\sqrt{\frac{\pi}{8}}+\frac{\sin\left(\frac{\sqrt{m}\eta_{-}^{2}t}{2\sqrt{J}}\right)}{\eta_{-}\sqrt{2\sqrt{m}t/\sqrt{J}}}-\frac{\cos\left(\frac{\sqrt{m}\eta_{-}^{2}t}{2\sqrt{J}}\right)}{4\left(\eta_{-}\sqrt{mt/2J}\right)^{3}}\right]\nonumber \\
		&-\sqrt{\frac{2\sqrt{m}}{\sqrt{J}t}}\sin\left(\frac{\sqrt{m}\eta_{-}^{2}t}{2\sqrt{J}}\right)\left[\sqrt{\frac{\pi}{8}}-\frac{\cos\left(\frac{\sqrt{m}\eta_{-}^{2}t}{2\sqrt{J}}\right)}{\eta_{-}\sqrt{2\sqrt{m}t/\sqrt{J}}}-\frac{\sin\left(\frac{\sqrt{m}\eta_{-}^{2}t}{2\sqrt{J}}\right)}{4\left(\eta_{-}\sqrt{\sqrt{m}t/2\sqrt{J}}\right)^{3}}\right]
	\end{align}
Eq.~\ref{Fresnel_asymp} further is simplified to give
\begin{equation}
	D_R \approx \frac{\sqrt{J}}{\sqrt{m}\eta_{-}^{3}t^{2}}
\end{equation}
Hence we see that, in the right sector, the OTOC has the following asymptotic form
\begin{equation}\label{OTOC_k=2_asymp}
	D(x, t)\approx
			\left|\frac{a}{2\pi}\left(\frac{\sqrt{J}}{\sqrt{m}\eta_{-}^{3}t^{2}}
			\right)			\right|^{2},\qquad
			\text{for $\eta_{-}> 0$ and large}
\end{equation}
We find that beyond the front, there is a power-law decay ($\propto \eta_{-}^{-6}$). We can see that this result (Eq.~\ref{OTOC_k=2_asymp}) agrees with the numerical simulations [Fig.~\ref{k2fig} (Right)] not only in terms of the exponent of the power law but also in terms of the coefficients given in Eq.~\ref{OTOC_k=2_asymp}.

%
%

\subsection{Non-integrable case, $k\ne 2$}

\subsubsection{$1< k<3$}

Now turning our attention to the next case, we try to analyse the asymptotic behaviour of 
\begin{equation}
D_\mathrm{R,L}(x,t)=\frac{a}{2\pi} \frac{B_k\left(\Delta_{\mp}\right)}{(3t\beta_k)^{1/k}},\,\,\text{ with }\,\Delta_{k, \mp}=\frac{x\mp \alpha t}{\big(3t\beta_k\big)^{1/k}}
\end{equation}
where we define a special function,
\begin{equation}
\label{Bfunc0}
B_k\left(y\right):=\int_{0}^{(3t\beta_k)^{1/k}\frac{\pi}{a}}ds\cos\left(ys+\frac{s^{k}}{3}\right)
\end{equation}
Again, we are interested in this analysis for large $N$ (equivalently, $a\rightarrow 0^{+}$) and large $\Delta_{k, \mp}$ limits of the system. The strategy would be to again look at the asymptotic form of Eq.~\ref{Bfunc0}. But in the absence of an analytical expression we resort to numerical methods. We find a power-law decay of the form
\begin{equation}\label{OTOC_hypo_test}
D(x, t)\propto|\Delta_{\pm}^{-(k+1)}|^{2}
\end{equation}
where $-/+$ corresponds to the right/left sectors, respectively. Therefore, the OTOC outside the light cone decays as a power law with exponent $2k+2$. 

\subsubsection{$3<k < 5$}
\begin{figure}[t!]
	\centering
	\includegraphics[width=0.32\textwidth,height=4.5cm]{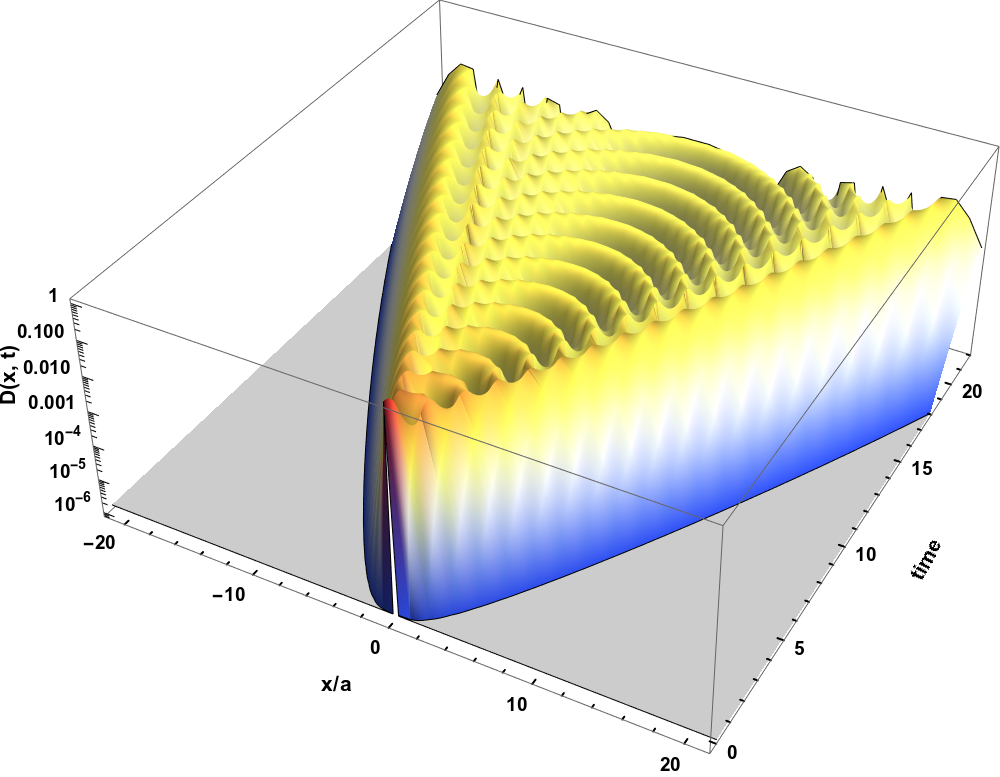}
	\includegraphics[width=0.32\textwidth,height=4.5cm]{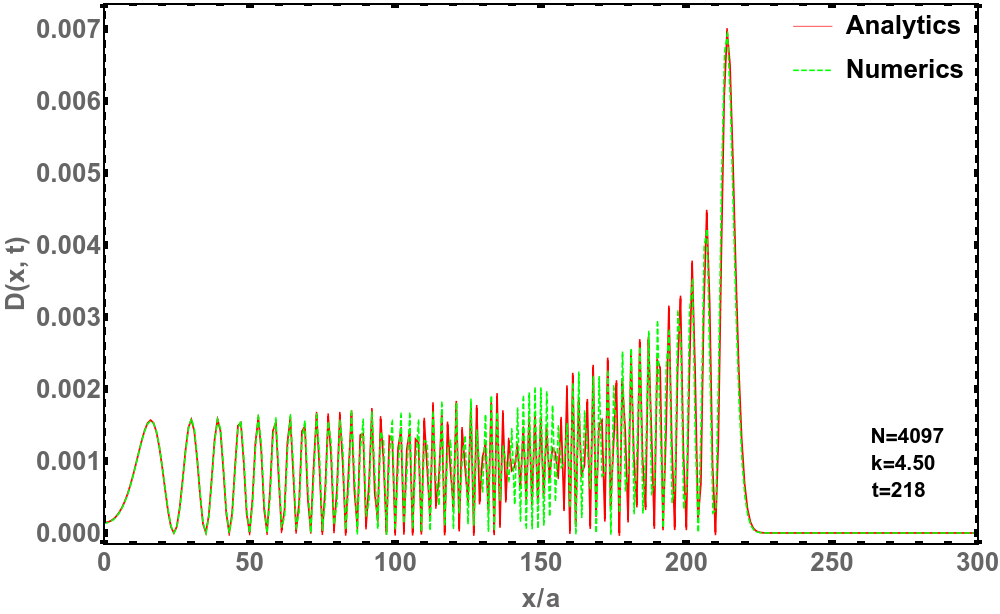}
	\includegraphics[width=0.32\textwidth, height=4.5cm]{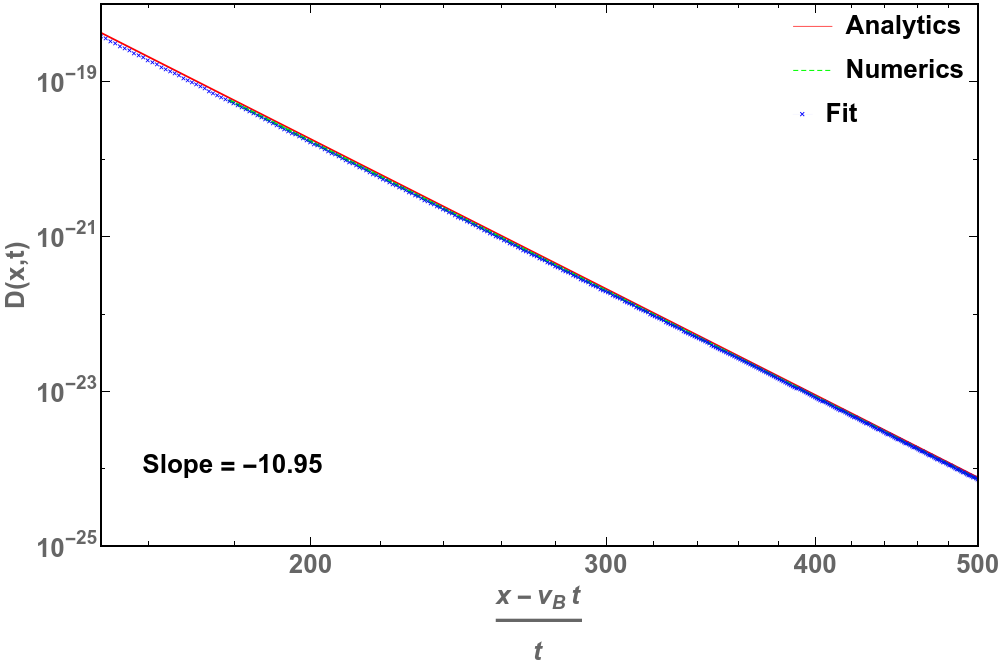}
	
	\caption{  (Left) Heat map of the OTOC, for $k=4.5$ and $N=65$, from direct numerical simulations. The time axis in in units of $a/v_{B}$. (Middle) Comparing OTOC from direct numerical simulation with the analytical expression at a time snapshot, $t\approx218$ (in units of $a/v_{B}$). Note that, for visualisation purpose,  we have plotted only the positive x-axis and the results are mirror symmetric on the other side. (Right) Power law decay of OTOC beyond the cone for $k=4.5$ and $N=4097$. The slope depicting the exponent of the power law. Here, $a \sim 0.1346$, $v_B \sim 466.514$}
	\label{k4p5fig}
	
\end{figure}

As we have discussed, just considering the first two terms in the expansion Eq.~\ref{Omega_expansion} does not capture the power-law decay of the OTOC. This is because we effectively end up with Airy function which has an exponentially decaying tail. So in order to observe the power-law nature of the OTOC, one has to consider the next order term in Eq.~\ref{Omega_expansion} as well, in addition to the first two terms.\\ 
The expectation is that once the Airy-like behaviour of the dominant term washes away, the power-law nature of the sub-dominant term will take over. This amounts to using Eq.~\ref{Omega_expansion} in Eq.~\ref{OTOC_integral}. We find that including the next order term $\sim q^k$ recovers the power-law from an Airy-like behaviour. For e.g., this claim about a power-law behaviour is verified, for $k=4.5$, by the direct numerical simulations and asymptotic analysis of the OTOC using the full dispersion relation (Eq.~\ref{PolyLog_DispRel}), as is seen in Fig.~\ref{k4p5fig} (Right). We find a decay $\propto \mid\Delta_{\pm}\mid^{-11}$ in this case.\\

Notice the drastic difference in the nature of the envelope of the OTOC profile as one goes from $k<2$ (peaking at the centre and then having a downward envelope) to $k=2$ (essentially flat) and $k>2$ (peaking at the edges and having an upward envelope). The integrable model ($k=2$) serves as a transition point for a change in the nature of propagation of the perturbations through the system.\\

\subsubsection{$k=3$}

The case of $k=3$, and more generally, odd integer values of $k$ is a little subtle from an analytical analysis stand point. As elaborated in Eq.~\ref{k to 3 limit}, the subtle cancellation of divergences arising due to the Riemann zeta function and the Gamma function gives rise to a logarithm term. This is a crucial term as it gives rise to the power-law decay of the OTOC. As is clear, in the absence of this term, we obtain Airy-like behaviour for $k=3$, due to the $q^{3}$ term in Eq.~\ref{Omega_expansion}. For $k=3$, using Eq.~\ref{Omega_expansion} in Eq.~\ref{OTOC_integral} produces a power-law decay of the OTOC $\propto \mid\Delta_{\pm}\mid^{-8}$.\\ 

These observations strongly suggest that the long-range nature of Riesz gas family of models is what gives rise to the power-law decay of the OTOC (for $k>1$). A particular term in the expansion of $\omega_k(q)$ at $O(q^k),\, \forall\,k>1$ explains the power-law. 

\section{Broyden–Fletcher–Goldfarb–Shanno (BFGS) algorithm}\label{BFGS}
We have implemented the BFGS algorithm \cite{fletcher1970new, broyden1970convergence} to find the global minimum energy configuration of our system (Eq.~\ref{Hamiltonian}) for system sizes upto $N=4097$. This is an iterative algorithm which can be used for non-linear, unconstrained optimisation problems. Below we provide the algorithm. 
Let $V(\{x_i\})$  (for e.g., Eq.~\ref{Hamiltonian}) be a scalar function on $\mathbb{R}^N$ which is to be minimised. $\nabla V(\{x_i\})$ is the corresponding gradient. $\textbf{M}$ is the Hessian matrix (for e.g., Eq.~\ref{eq:hessian}).
\begin{itemize}
	\item[1] Make an initial guess for the minimising configuration $\textbf{x}_0=\{x^0_i\}$. Using this compute $\textbf{M}_0=\textbf{M}(\textbf{x}_0)$ and $\nabla V_0=\nabla V(\textbf{x}_0)$.
	
	\item[2] Find the minimising direction by solving: $\textbf{M}_n \textbf{p}_n = \nabla V_n$. Here $\textbf{p}_n$ is the minimisation direction at the $n^{th}$ iterative step. Therefore, we solve the inverse problem $\textbf{p}_n = \textbf{M}_n^{-1}  \nabla V_n$. 
	
	\item[3] Now, we try to find the optimum step size $\alpha_n$ to take in the minimisation direction. One could use the back-tracking line-search method for this purpose. We start with a sufficiently large initial step size $\alpha_0$ and iteratively reduce it, i.e., $\alpha_{j+1}=\eta \alpha_{j}$ where $\eta\in(0,1)$ is a control parameter. This is done until the Armijo-Goldstein condition is satisfied, namely, $V(\textbf{x}_n)-V(\textbf{x}_n+\alpha_j\textbf{p}_n)>\alpha_j \gamma$, where $\gamma=-c \delta$ and $\delta=\nabla V.\textbf{p}_n$. Here, similar to $\eta$,  $c\in (0,1)$ is another control parameter. So, we finally find $\alpha_n$ such that it is the largest of the set $\alpha_j's$ until  the Armijo-Goldstein condition is still satisfied. 
	
	\item [4] Updating procedure:
	\begin{itemize}
		\item[o] $\textbf{x}_{n+1}=\textbf{x}_n+\alpha_n\textbf{p}_n$.
		\item[o] $\textbf{s}_n = \alpha_n\textbf{p}_n$
		\item [o]$\textbf{y}_n=\nabla V_{n+1}-\nabla V_n$
		\item[o] $\textbf{M}_{n+1}=\textbf{M}_{n}+\frac{\textbf{y}_n\textbf{y}_n^T}{\textbf{y}_n^T\textbf{s}_n}-\frac{\textbf{M}_{n}\textbf{s}_n\textbf{s}_n^T\textbf{M}_{n}^T}{\textbf{s}_n^T\textbf{M}_{n}\textbf{s}_n}$
	\end{itemize} 
	\item[5] Repeat steps $2-4$.
\end{itemize}


\bibliography{references.bib}